\documentclass[acmsmall]{acmart}
\AtBeginDocument{%
  }

\usepackage{multirow}
\usepackage{array}
\usepackage[suppress]{color-edits}
\usepackage{xcolor}
\usepackage{microtype}
\usepackage{hyperref}
\usepackage{url}
\usepackage{booktabs}
\usepackage{graphicx}
\usepackage{color-edits}
\usepackage{booktabs, multirow}
\usepackage{pdflscape}
\usepackage{amsmath, amsfonts}
\usepackage{bbm}
\usepackage{nicefrac}       
\usepackage{microtype}      
\usepackage{pdfpages}
\usepackage{makecell}
\usepackage{placeins}
\usepackage{afterpage}

\addauthor[Stevie]{sc}{brown}
\addauthor[Devansh]{ds}{red}
\addauthor[Ken]{kh}{violet}
\addauthor[Author]{lg}{blue}

\setcopyright{cc}
\setcctype{by}
\acmJournal{PACMHCI}
\acmYear{2025} \acmVolume{9} \acmNumber{7} \acmArticle{CSCW447} \acmMonth{11} \acmPrice{}\acmDOI{10.1145/3757628}
\acmISBN{978-1-4503-XXXX-X/2018/06}

\begin{document}

\title[Measurement as Bricolage]{Measurement as Bricolage: Examining How Data Scientists Construct Target Variables for Predictive Modeling Tasks}


\author{Luke Guerdan$^\dagger$}
\email{lguerdan@cs.cmu.edu}
\affiliation{%
  \institution{Carnegie Mellon University}
  \city{Pittsburgh}
  \state{PA}
  \country{USA}}

\author{Devansh Saxena$^\dagger$}
\affiliation{%
  \institution{University of Wisconsin-Madison}
  \city{Madison}
  \state{WI}
  \country{USA}}

\author{Stevie Chancellor$^*$}
\affiliation{%
  \institution{University of Minnesota}
  \city{Minneapolis}
  \state{MN}
  \country{USA}}

\author{Zhiwei Steven Wu$^*$}
\affiliation{%
  \institution{Carnegie Mellon University}
  \city{Pittsburgh}
  \state{PA}
  \country{USA}}

\author{Kenneth Holstein$^*$}
\affiliation{%
  \institution{Carnegie Mellon University}
  \city{Pittsburgh}
  \state{PA}
  \country{USA}}

\thanks{$\dagger$ Co-first authors contributed equally to this research. $^*$Co-senior authors contributed equally to this research.}

\renewcommand{\shortauthors}{Luke Guerdan, Devansh Saxena, Stevie Chancellor, Zhiwei Steven Wu, \& Kenneth Holstein}

\begin{abstract}
Data scientists often formulate predictive modeling tasks involving fuzzy, hard-to-define concepts, such as the ``authenticity'' of student writing or the ``healthcare need'' of a patient. Yet the process by which data scientists translate fuzzy concepts into a concrete, proxy target variable remains poorly understood. We interview fifteen data scientists in education (N=8) and healthcare (N=7) to understand how they construct target variables for predictive modeling tasks. Our findings suggest that data scientists construct target variables through a \textit{bricolage} process, in which they use creative and pragmatic approaches to make do with the limited data at hand. Data scientists attempt to satisfy five major criteria for a target variable through bricolage: validity, simplicity, predictability, portability, and resource requirements. To achieve this, data scientists adaptively apply \textit{problem (re)formulation strategies}, such as \textit{swapping} out one candidate target variable for another when the first fails to meet certain criteria (e.g., predictability), or \textit{composing} multiple outcomes into a single target variable to capture a more holistic set of modeling objectives. Based on our findings, we present opportunities for future HCI, CSCW, and ML research to better support the art and science of target variable construction.

\end{abstract}



\received{October 2024}
\received[revised]{April 2025}
\received[accepted]{August 2025}

\maketitle

\section{Introduction}

\begin{quote}
\textit{``Problem-solvers who do not proceed from top-down design but are arranging and rearranging a set of well-known materials can be said to be practicing bricolage. They tend to try one thing, step back, reconsider, and try another''} (Turkle, p. 51).
\end{quote}

Data scientists grapple with subtle but important considerations when translating organizational goals into a formal modeling task. For example, consider a data science team developing a model to help physicians decide which patients have sufficient ``healthcare need’’ for enrollment in a high-risk care management program (e.g., \citep{obermeyer2019dissecting}). While developing a model, data scientists may work with clinicians to understand how to translate risk predictions to actionable interventions --- for example, by selecting an appropriate risk cut-off for enrollment recommendations. Data scientists may also carefully consider whether the patient populations represented in their training data adequately match the characteristics of patients encountered when the model is deployed. The process of \textit{problem formulation} describes how data scientists navigate such translational concerns as they map a high-level goal into a tractable computational problem \citep{passi2019problem}.

\begin{figure}[h]
    \centering
    \includegraphics[width=.27\linewidth, trim=7mm 6mm 8mm 6mm, clip]{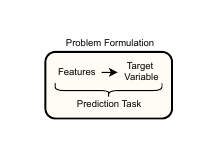}
    \caption{An illustration of the relationship between a target variable, prediction task, and problem formulation. A \textit{prediction task} describes a mapping from \textit{features} to a \textit{target variable}. A \textit{problem formulation} connects a prediction task to high-level modeling goals, including factors such as how predictions are converted to decisions.\looseness=-1}\label{fig:formulation_onion}
\end{figure}

One critical aspect of problem formulation is the process of \textit{target variable construction} (Figure \ref{fig:formulation_onion}). Many outcomes of interest to data scientists --- such as the ``authenticity'' of student writing or the ``healthcare need'' of patients --- are latent, theoretical constructs that are not directly observable. To develop a model targeting these constructs, data scientists must construct a \textit{proxy variable} based on outcomes that are readily available. For example, a data scientist might choose to predict patients' cost of medical care as a proxy for their ``healthcare need'' \citep{obermeyer2019dissecting}. Traditionally, measurement theory assumes that analysts follow a top-down workflow while mapping unobserved constructs to measurements in data \citep{jacobs2021measurement}. For example, a learning scientist might identify a latent construct they would like to measure among students (e.g., ``math proficiency'') and design an operationalization (e.g., a test) \textit{before} collecting data (e.g., student test scores). Disciplines in the quantitative social sciences have developed a rich set of guidelines (e.g., \citep{van2016pre, miller2002handbook}) and modeling tools (e.g., \citep{everett2013introduction, singh2007quantitative}) to help analysts implement this top-down measurement workflow.\looseness=-1 

\textbf{However, the traditional top-down conceptualization of measurement is in tension with our understanding of data science as a \textit{bottom-up} process constrained by \textit{existing} data} \citep{passi2019problem, kross2021orienting}. \citet{passi2019problem} observe that problem formulation in data science is an iterative process, where data scientists progressively reconcile their high-level objectives with low-level data and modeling constraints. As a result, while a traditional top-down measurement approach assumes that an analyst has the resources to collect data tailor-made to the goals of a study, data scientists are often forced to re-purpose the limited data that happens to be available. This disconnect creates a significant challenge: though many flaws in data science projects can be traced back to target variable construction (e.g., \citep{turque2012creative, amrein2014rethinking, mullainathan2021inequity, kawakami2022improving, cheng2022child}), existing tools, frameworks, and pedagogical resources reflect a top-down planning approach towards measurement, rarely accounting for the bottom-up, resource-constrained reality of data science practice.

To understand data scientists' current practices and challenges in constructing target variables---and to identify opportunities to better support target variable construction in practice---in this work, we conducted semi-structured interviews with fifteen data scientists from education (N=8) and healthcare (N=7). We asked data scientists to walk us through how they navigated target variable construction in prior modeling projects. Participants also engaged in a hypothetical target variable construction task, allowing us to observe their reasoning in a new modeling context. Our goal in these interviews was to understand how data scientists intertwine two conceptually distinct activities, measurement of latent constructs and prediction of target variables, during problem formulation \citep{mussgnug2022predictive}. While prior empirical studies of data science practice have often centered prediction (e.g., \citep{alspaugh2018futzing, wongsuphasawat2019goals, kandel2012enterprise}), we broaden our scope of inquiry to investigate the interplay between measurement validity and predictive utility in problem formulation.

We find that data scientists construct target variables by \textbf{leveraging creative and pragmatic approaches to make do with the limited data they have at hand}. Data scientists craft their formulation through a process of iterative \textit{dialogue} with their data, tweaking and refining their target variable until they conclude that it provides a sufficient operationalization for the unobserved construct of interest. To understand data scientists’ target variable construction practices, we draw upon the concept of \textit{bricolage}, originally introduced by Claude Lévi-Strauss to contrast the rigid top-down methodology of Western science with the more tangible ``science of the concrete’’ practiced in non-Western societies \citep{lvi1966savage}. Rather than implementing abstract concepts through top-down design, the bricoleur solves problems by arranging and re-arranging well-known materials in a bottom-up process. Like bricoleurs, data scientists in our study often acknowledged imperfection in their improvisational approach to measurement. Yet they \textit{also} identified innovative solutions that would be difficult to anticipate through a process of top-down design.


Based on our findings, \textbf{we synthesize a new process model that better reflects data scientists' bricolage approach towards measurement} in problem formulation (Figure \ref{fig:main_findings}). Data scientists balance the validity of their formulation with other criteria such as predictive performance, simplicity, portability, and resource requirements. Data scientists narrow in on a suitable formulation by drawing upon a rich repertoire of problem (re)formulation strategies. Data scientists apply a strategy, take a step back to assess their progress along each criterion, and then apply a \textit{new} strategy until they believe all evaluation criteria are satisfied. Data scientists sometimes discontinue their projects if they conclude that no strategies are sufficient given the resources at hand.\looseness=-1

Our characterization of target variable construction as bricolage \textbf{offers an important counterbalance to nascent views on measurement of AI systems}. While recent work correctly identifies measurement problems as the source of many AI failures \citep{coston2023validity, kawakami2024situate, guerdan2023ground, jacobs2021measurement, raji2022fallacy, liu2024ecbd, raji2021ai, alaa2025medical}, proposed solutions often assume or reinforce a top-down, idealized measurement process. Our findings suggest that in order for established measurement principles to have impact in practice, there is a need for thoughtful reconciliation with the resource-constrained, bottom-up reality of data science work. By documenting how measurement and prediction intertwine in data scientists' bricolage practices, we contribute a more grounded understanding of how measurement actually occurs in the wild, providing a foundation for developing more effective interventions. While discussing implications of our findings, \textbf{we present new opportunities for future CSCW, HCI, and ML research to develop tools that better support data scientists’ bricolage practices}. 

In summary, our contributions to the CSCW literature are: 

\vspace{-\topsep}
\begin{itemize}
    \itemsep0em 
    \item A novel characterization of measurement in data science as a \textit{bricolage practice}. We illustrate how data scientists adaptively balance multiple evaluation criteria by creatively combining existing data under resource constraints.
    \item Empirical findings from interviews with fifteen data scientists in healthcare and education, synthesized into a process model that reveals how practitioners weigh competing criteria ($\S$ \ref{subsec:tradeoffs}), apply five strategic approaches to (re)formulation ($\S$ \ref{subsec:strategies}), and evaluate validity through both theory-driven and data-driven practices ($\S$ \ref{subsec:evaluating_construct_validity}).
    \item Design implications for how the CSCW, HCI, and ML research communities might better scaffold data scientists' bricolage practices going forward. These recommendations explore how we might enhance the validity of data scientists' resulting formulations while also supporting the flexibility that makes bricolage effective ($\S$ \ref{sec:discussion}).  
\end{itemize}
\vspace{-\topsep}

\section{Background and Related Work}

In this section, we begin by introducing Claude Lévi-Strauss' theory of bricolage. We then describe existing Science and Technology Studies (STS), HCI, and CSCW research that has investigated questions related to categorization, measurement, and target variable construction in data science practice. We conclude by providing an overview of key concepts from measurement theory in the quantitative social sciences which we build upon in this work.

\subsection{Bricolage and the Science of the Concrete}\label{subsec:bricolage_background}

\textit{Bricolage}, defined by Claude Lévi-Strauss as ``making the most of available resources''\citep{lvi1966savage}, describes how actors (i.e., \textit{bricoleurs}) construct artifacts in creative and adaptive ways. Bricolage is central to what Lévi-Strauss calls \textit{the science of the concrete}, which contrasts with modern scientific thought by focusing on the materials directly at hand rather than abstract principles. Bricolage contains three elements: repertoire, dialogue, outcome \citep{duymedjian2010towards}. \textit{Repertoire} describes the resources---physical artifacts, ideas, or skills---that the bricoleur brings to bear on a problem. Whereas the engineer or scientist acquires resources tailor-made to specification, the bricoleur assembles ``odds and ends'' that might be useful in the future. Then, when faced with a new problem, the bricoleur engages in \textit{dialogue} with these materials---a process of arranging and re-arranging materials in her repertoire to find a workable solution. While engaging in dialogue, the bricoleur adopts old objects for new purposes and combines concepts with materials in novel ways. Finally, \textit{outcome} describes the artifact resulting from the bricolage process. While bricolage entails a tolerance for imperfection, bricoleurs sometimes identify ``brilliant unforeseen results''  \citep{lvi1966savage} as part of their process. 

Prior research has used Lévi-Strauss' framework to describe many forms of knowledge work as bricolage (e.g., entrepreneurship \citep{mateus2024bricolage, duymedjian2010towards}, design \citep{buscher2001landscapes,louridas1999design,vallgaarda2015interaction}, management \citep{duymedjian2010towards}, and programming \citep{turkle2011life}). For instance, \citet{baker2005creating} use bricolage to explain how entrepreneurs build business ventures under resource constraints. They observe that entrepreneurial success hinges on the ability of firms to combine their limited \textit{existing} resources in creative, adaptive ways. 

In her book \textit{Life on the Screen}, Sherry Turkle uses bricolage to describe how programmers write computer code. Through the mid-1980s, the canonical coding approach taught in classrooms was \textit{structured programming}---a linear, top-down process that proceeds by systematically mapping abstract plans into procedural instructions. Yet, Turkle observed that programmers---novices and experts alike---often engaged in a more unstructured \textit{dialogue} with their code while solving real-world problems. Drawing an analogy between the programmer and bricoleur, she explains:

\begin{quote}
\textit{``In the context of programming, the bricoleur's work is marked by a desire to play with lines of computer code, to move them around almost as if they were material things---notes on a score, elements of a collage, words on a page''} (Turkle, p. 51-52)
\end{quote}

Turkle describes this style of tinkering as important for developing an intuition for the complexity of computer programs. Like a painter who steps back to inspect a canvas between brush strokes, programmers try one approach, re-evaluate, and then try another. This process often leads programmers to more elegant solutions than could be imagined through structured planning. 

In this work, we draw a novel connection between \textit{data science work} and bricolage. Much like an entrepreneur, programmer, or painter, data scientists construct target variables by combining \textit{existing} resources in creative, adaptive ways. Instead of instantiating abstract plans through top-down design---akin to the structured programming approach taught in classrooms---data scientists engage in more direct \textit{dialogue} with their data and modeling tools, making do with the materials at hand. And analogous to past research investigating how to support such exploratory and improvisational coding practices~\cite{kery2017exploring}, this has implications for the design of tools to support target variable construction in practice. We highlight connections between target variable construction and bricolage throughout the findings ($\S$ \ref{sec:findings}) and describe implications for tool design in the discussion ($\S$ \ref{sec:discussion}).\looseness=-1 

\subsection{Tracing the Social Construction of Target Variables}

Target variable construction also relates to a rich body of Science and Technology Studies and philosophy of science literature that investigates the social processes that give rise to categories in data \citep{latour2013laboratory, bowker2000sorting, pine2015politics, harvey2024cadaver, porter1996trust, crosby1997measure, hacking1999social, mol2002body, suchman1993categories}.  \citet{latour2013laboratory} describe the social construction of scientific facts within laboratories, highlighting that facts are not merely discovered but actively produced through a complex interplay of social and technical factors. Likewise, \citet{bowker2000sorting} examines how the International Classification of Diseases (ICD), shape and are shaped by the social and political contexts in which they are embedded. Building on Bowker's work, recent studies have also explored the construction of racial and ethnic categories in AI training data (e.g., \citep{abdu2023empirical, mickel2024racial}) and charted measurement assumptions underpinning motion capture technologies \citep{harvey2024cadaver}.

Most related to our study, \citet{muller2021designing} interviewed fifteen data scientists to investigate the \textit{labeling} practices of data science teams. \citet{muller2021designing} identified three approaches that teams used while defining labels: \textit{principled}, \textit{iterative}, and \textit{improvisational}. Teams engaging in principled design followed a top-down planning approach, systematically specifying annotation guidelines before data labeling. Teams leveraging iterative design took a bottom-up approach, incorporating opportunities for annotation protocol refinement throughout the labeling process. Teams engaging in improvisational design viewed data labeling as an open-ended, exploitative process. Our study investigates a distinct question of how data science teams construct target variables for \textit{predictive modeling tasks}. Rather than studying how data science teams design ``ground truth'' labels anew, we examine how data scientists connect their \textit{existing} data to their high-level modeling objectives.\looseness=-1

\subsection{Empirical Studies of Data Science Practice}\label{subsec:empirical_studies}

Our work also builds upon empirical studies investigating how data scientists conduct their day-to-day work. Many studies have examined activities in the technical ``inner loop'' \citep{kross2021orienting} of data science, such as exploratory data analysis \citep{alspaugh2018futzing, wongsuphasawat2019goals}, feature engineering, \citep{muller2019data}, and visualization, \citep{kandel2012enterprise}. Other studies have investigated ``outer loop'' activities performed by data scientists, including collaborating with other stakeholders \citep{kross2021orienting, passi2018trust, nahar2022collaboration, mao2019data, kim2016emerging, zhang2020data, hou2017hacking} and addressing ethical concerns \citep{holstein2019improving, deng2022exploring}. Prior work has characterized inner and outer loop activities as iterative and interconnected \citep{kross2021orienting, passi2019problem}. In particular, based on interviews with ten professionals spanning industry and academia, \citet{kross2021orienting} observed that data scientists iteratively ``framed the problem'' (i.e., an outer loop activity) then returned to their data to perform analyses (i.e., inner loop activities). 

\textbf{However, we currently have an incomplete understanding of \textit{how} data scientists construct target variables} while formulating prediction tasks. To date, \citet{passi2019problem} provide the most detailed account of how problem formulation unfolds in real-world modeling contexts. Over the course of an eighteen-month ethnographic study of a corporate data science team, \citet{passi2019problem} observed that data scientists repeatedly (re)formulated their prediction task as they discovered new information about data availability constraints and the predictability of alternative outcome variables. Based on their findings, \citet{passi2019problem} describe problem formulation as a \textit{negotiated translation} between high-level objectives and a concrete prediction task. Our work builds upon this finding to identify the more general evaluation desiderata and (re)formulation strategies driving the problem formulation process. We develop a process model through analysis of our findings and use this model to identify targeted opportunities to better support data scientists' problem formulation practices going forward. For example, when viewed through the lens of our findings, the team studied by \citet{passi2019problem} used \textit{one of five} (re)formulation strategies (i.e., \textit{swapping}) to change outcome variables when the first proved infeasible due to data availability constraints. We present design opportunities to help data scientists more effectively apply swapping and other (re)formulation strategies in Section \ref{sec:discussion}.


\subsection{Toolkits Supporting Target Variable Construction}

The limitations of data science toolkits and analysis methods have been documented across empirical studies examining data science practices \citep{jun2022hypothesis}, software repositories \citep{hoess2025does}, and ML-based sciences \citep{kapoor2023leakage}. Jun et al. reviewed 20 commonly-used data analysis tools and found that the ecosystem of tools supports vastly different model implementations, even when using the same underlying mathematical equation \citep{jun2022hypothesis}. Although these tools support nuanced model implementations, their low-level technical abstractions make it challenging for analysts to engage in the high-level conceptual reasoning required for hypothesis formalization \citep{jun2022hypothesis}.  As a result, modeling assumptions and interpretations of data can impact the validity of results \citep{hoess2025does, jun2022hypothesis, jun2022tisane, kapoor2023leakage}. Research examining software engineering practices in code repositories, such as GitHub, further demonstrates how the choice of analysis tools can lead to substantial differences in simple metrics (e.g., commit counts and developer contributions) \citep{hoess2025does, harding2021diffdelta, lefever2021lack}. This further poses threats to validity of findings derived from one analysis tool. Given this gap between technical abstractions and conceptual reasoning—and the embedded assumptions inherent in machine learning methods like feature engineering—it is not surprising that ML-based sciences continue to face a reproducibility crisis \citep{kapoor2023leakage}. 



To mitigate these issues, the HCI, ML, and FAccT research communities have developed a range of toolkits to intervene in data scientists’ inner and outer loop workflows. For instance, researchers have developed systems facilitating data collection and configuration (e.g., \citep{bhattacharya2024exmos, guo2011proactive, kandel2011wrangler, rattenbury2017principles}), exploratory data analysis (e.g., \citep{wongsuphasawat2015voyager, kery2018story}), model training (e.g., \citep{pang2020deep, dingen2018regressionexplorer}), caching intermediate states during data science experiments (e.g., \citep{schubotz2022caching}), and performance assessment (e.g., \citep{cabrera2023zeno, yan2021tessera}). Many tools have also been developed to help data scientists explain their models (e.g., \citep{ribeiro2018anchors, ribeiro2016should, lundberg2017unified}) or assess fairness-related concerns (e.g., \citep{tramer2017fairtest, bellamy2019ai, weerts2023fairlearn, saleiro2018aequitas, liu2024faircompass, cabrera2019fairvis, ahn2019fairsight, wexler2019if, dingen2018regressionexplorer}). In the FAccT community, researchers have also developed responsible AI checklists and guidelines to help data scientists assess and document components of their AI system (e.g., dataset information \citep{gebru2021datasheets, pushkarna2022data}, intended use cases \citep{mitchell2019model}). To our knowledge, \citet{gala2024fairtargetsim} and \citet{sivaraman2025tempo} propose the only systems specifically designed to support target variable construction. Instead, existing toolkits are often designed to support data scientists' practices only \textit{after} a target variable has been specified. This presents a pressing gap in data science support at a critical point in their workflows. 

However, prior human--computer interaction research suggests that new toolkits may prove ineffective if they are not designed around the \textit{existing} practices of data scientists. For instance, past efforts to improve programming tools without a human-centered assessment of programmers' needs yielded ineffective languages and systems \citep{myers2016programmers, hazzan2004human, seffah2005human, ko2011state}. Similarly, the utility of ML fairness toolkits was found to be limited in practice as they were designed for isolated use by technical roles, while actual fairness work often relied on deep cross-functional collaborations~\citep{deng2022exploring}. Our work aims to help the research community avoid similar challenges while developing tools to support data scientists' problem formulation processes. In particular, we identify opportunities for the HCI, CSCW, and ML research communities to develop tools that help data scientists more carefully weigh trade-offs across multiple evaluation criteria ($\S$ \ref{subsection:supporting_validity_tradeoffs}), select ($\S$ \ref{subsection:identifying_strategies}) and apply ($\S$ \ref{subsection:applying_strategies}) strategies, and evaluate the validity of their problem formulation more effectively ($\S$ \ref{subsection:supporting_validity_evaluations}).\looseness=-1

   
\subsection{Measurement and Validity in AI}

A growing line of research traces AI system failures to \textit{validity} pitfalls \citep{jacobs2021measurement, milli2021optimizing, stray2021you, guerdan2023ground, goyal2022your, obermeyer2019dissecting, mullainathan2021inequity, coston2023validity}. In the quantitative social sciences, the validity of a measurement instrument is established via a multifaceted assessment along several dimensions. \textit{Construct validity} describes the extent to which a measurement instrument accurately reflects the theoretical construct it purports to measure \citep{jacobs2021measurement}. In predictive modeling, construct validity is most often discussed in connection to whether the target variable predicted by a model reflects a theoretical construct of interest to model developers \citep{guerdan2023ground, coston2023validity}. \footnote {\citet{jacobs2021measurement} provide a detailed review of sub-dimensions of construct validity as they pertain to measuring fairness properties of algorithmic systems.} \textit{Internal} and \textit{external} validity are also important dimensions in experimental research. \textit{Internal validity} describes the extent to which an experimental study can confidently establish a causal relationship between independent and dependent variables \citep{gergle2014experimental}. In predictive modeling, it relates to the existence of a defensible causal relationship between features and a target variable \citep{coston2023validity}. \textit{External validity}, on the other hand, measures how well an analysis can be generalized beyond the specific conditions of the study. This dimension has been linked to the extent to which a predictive model's in-distribution performance generalizes across real-world deployment contexts \citep{coston2023validity}. 

While these validity dimensions articulate helpful conceptual desiderata for a predictive model, they provide a limited picture of \textit{how} practitioners actually navigate measurement decisions in their day-to-day work. \citet{mussgnug2022predictive} argue that current practice is characterized by a ``predictive reframing'', in which data scientists cast a measurement problem (inferring existing but unobserved quantities) as a prediction task. This reframing shifts the epistemic aim from measuring a concept to predicting a given measurement of that concept. For example, when data scientists attempt to assess poverty levels using satellite imagery, they often frame their task as ``predicting poverty'' rather than ``measuring poverty.'' This practice centers the evaluation of a machine learning model around narrow predictive performance characteristics while sidestepping deeper questions of validity. 

We expand upon \citet{mussgnug2022predictive}'s argument by characterizing how data scientists weigh validity and predictive performance during problem formulation. In support of \citet{mussgnug2022predictive}'s argument, we find that many prediction tasks pursued by data scientists are indeed measurement problems. For example, some participants in our study used supervised learning models to measure concepts recorded at future points in time --- e.g., ``10-year cardiovascular risk.'' Others used supervised learning models as measurement instruments while measuring a concept at a current point of time --- e.g., to score the ``authenticity'' of a student's writing. In both cases, data scientists heavily anchored on predictive performance while evaluating their problem formulation. However, counter to \citet{mussgnug2022predictive}'s argument, we find that data scientists \textit{do} also engage with validity concepts during problem formulation, albeit with a more pragmatic, situated approach that lacks formal terminology. We uncover how data scientists evaluate validity in our findings ($\S$ \ref{sec:findings}), and unpack implications for data science pedagogy ($\S$ \ref{subsection:pedagogy}) and tooling support ($\S$ \ref{subsection:supporting_validity_evaluations}) in the discussion.



\section{Methods}

In this study, we interviewed fifteen data scientists from the education (N=8) and healthcare (N=7) sectors who had prior experience with developing and evaluating predictive models. We conducted one-hour semi-structured interviews to understand participants’ current practices, perceived challenges, and future opportunities for target variable construction in predictive modeling tasks. 

\subsection{Study Design}

\subsubsection{Directed Storytelling} During the first interview segment, we used a directed storytelling approach~\cite{evenson2016case}, inviting data scientists to reflect on a time in the past when they developed and evaluated a predictive model. To ground our discussions, we first asked participants to share a few concrete experiences where they had developed and/or evaluated predictive models as part of their role. Building on these experiences, the interviewer then narrowed in on one of the projects mentioned and asked participants the following follow-up questions, in a semi-structured fashion:
\begin{enumerate}
    \item Were there ever times when you felt uncertain about your choice of target variable?
    \item Could you walk me through how your team went about picking a target variable? 
    \item How did you know whether your team selected the “right” target variable? 
\end{enumerate}

We also asked participants additional follow-up questions when these were not already covered by participants during their earlier responses, such as measurement guidelines and best practices they consulted throughout their project. On average, the first segment lasted 30-40 minutes.

\begin{table}
  \begin{tabular}{>{\centering\arraybackslash} m{1cm} >{\centering\arraybackslash} m{1.5cm} >{\centering\arraybackslash} m{2cm} >{\centering\arraybackslash} m{4.5cm} >{\centering\arraybackslash} m{3cm} }
     \toprule
     \makecell{\vspace{-2mm}\centering PID} & \makecell{\vspace{-2mm}\centering Domain} & \makecell{Postgrad\\Experience}  & \makecell{Current\\Primary Role} & \makecell{Current\\Affiliation} \\ 
     \midrule
     P1 &  \multirow{8}{*}{Education} & 15 years & Board Member & Industry  \\ 
     P2 &   & 7 years & Senior Analyst, Ph.D. Student & Industry/Academic  \\
     P3 &  & 5 years & Postdoc & Academic  \\
     P4 &  & 2 years & Ph.D. Student & Academic  \\
     P5 &  & 2 years & Ph.D. Student & Academic  \\
     P6 &  & 11 years & VP of Data Science & Industry  \\
     P7 &  & 10 years & Data Scientist & Industry  \\
     P8 &  & 11 years & Senior Research Scientist & Industry  \\
     \hline
     P9 &    & 7 years & Senior Research Scientist & Industry  \\
     P10 &  & 6 years & Research Associate & Academic  \\
     P11 &   & 4 years & Senior Data Scientist & Industry  \\
     P12 &  Healthcare & 11 years & Senior Data Science Analyst & Industry  \\
     P13 &   & 8 years & Senior Data Scientist & Industry  \\
     P14 &   & 8 years & PostDoc & Academic  \\
     P15 &   & 10 years & Assistant Professor & Academic  \\
     \bottomrule
    \end{tabular}
    \caption{Demographic overview of study participants: data scientists' experience, roles, and affiliations.}\label{tab:participants}
    \vspace{-2mm}
\end{table}

\subsubsection{Vignette-Based Problem Formulation Task} During the second segment, we asked participants to reason about a hypothetical problem formulation as we walked them through a vignette. We aimed to better understand how participants reason about measurement decisions on-the-fly, while potentially surfacing challenges they might not have recalled while reflecting on their prior experiences. We asked participants with a healthcare background to imagine they were developing a model to predict patients with high future ``medical acuity,'' and we asked participants with an education background to imagine they were developing a model to predict which students were ``academically at-risk'' (see Appendix \ref{appendix:design_probe_details}).\footnote{P1 was shown the healthcare vignette due to data availability constraints encountered at early phases of the interview schedule. During the interview, P1 signaled places where their interpretation of the vignette was informed by their domain experience. We took this into consideration during analysis.} We provided participants a set of features and outcome variables and instructed them to imagine that these were the data available for model development.

During the vignette, we asked participants to reflect on which outcomes might serve as a ``better or worse choice'' of prediction target given the stated modeling goals. After learning how participants initially approached this task, we then presented a series of model evaluation plots indicating the performance of predictive models targeting each outcome (see Appendix \ref{appendix:design_probe_details}). After showing each plot, we paused and asked participants to again reflect on which outcomes might serve as better or worse choices for the stated modeling goals. We also asked participants to share any additional information that they would ideally like to have, to better inform their thought process. As participants engaged with the task over subsequent interviews, we refined the specificity of the scenario description and accompanying evaluation plots to further probe issues that surfaced in prior interviews. This second segment lasted 20-30 minutes of the interview on average.

\subsection{Participants}

Participants were recruited via purposive sampling techniques \citep{heckathorn2011comment} from authors' professional networks (Table \ref{tab:participants}). To ensure that participants had a baseline level of domain expertise, we required that data scientists had domain-specific graduate-level training related to healthcare or education (e.g., biostatistics, learning sciences) or at least one year of full-time industry experience in a relevant domain to be eligible to participate. We further required that participants had prior experience developing and evaluating predictive models as part of their role. Our participants had a range of post-graduate experience at the time of the interview, spanning from 3-6 months to 15 years. At the time of the interview, participants had a variety of primary roles (e.g., Data Scientist, Assistant Professor, Board Member, PostDoc) from both academic and industry affiliations. As a result, our sample consists of data scientists with a baseline level of domain knowledge and a range of technical experience levels. This study was approved by the University Institutional Review Board and all participants were compensated with a \$35 Amazon gift card for completing the study.\looseness=-1

\subsection{Data and Analysis}

One of the authors conducted interviews remotely via Zoom between September and December 2023. In total, we recorded and transcribed fourteen hours and thirty minutes of audio after collecting verbal and written consent from all participants. Two authors independently performed open coding on transcripts and met regularly to discuss and reconcile interpretation differences. After interviewing fifteen participants, we observed a common set of themes recurring across interviews and took this as an indicator that we had reached saturation \citep{saunders2018saturation}. We adopted a reflexive thematic analysis approach \citep{braun2006using, clarke2017thematic} to group codes into successively higher themes, informed by synchronous discussions among all of the authors. Our process gave rise to sixteen high-level themes, organized into three categories: (1) \textit{factors} driving the target variable construction process, which we discuss in $\S$\ref{subsec:tradeoffs}; (2) data scientists’ \textit{model development} and \textit{evaluation} practices, which we discuss in $\S$\ref{subsec:strategies} and $\S$\ref{subsec:evaluating_construct_validity}; and (3) data scientists’ \textit{orientation} towards the problem formulation process, which we discuss throughout the findings, when drawing connections to the concept of bricolage. 

\lgedit{
\subsection{Positionality}

Our positionality as a research team is shaped by our backgrounds spanning human-computer interaction, cognitive psychology, machine learning, and design. Several team members have professional experience as data scientists, while others have training in educational analytics and behavioral health. This interdisciplinary background helped us recruit participants and build rapport during interviews. During analysis, our team's quantitative social sciences background helped us connect the practices we documented to established ideas in validity theory. Our interdisciplinary perspective also directed our focus toward tensions between pragmatic data science realities—which we personally encountered—and formal validity concepts from academic research. Team members with design experience had previously encountered Claude Lévi-Strauss' concept of bricolage, which became our central theoretical framework. The pragmatic orientation in our findings emerged from combining academic social science training with real-world data science experience.}

\begin{figure}
    \centering
    \includegraphics[width=\linewidth, trim=3mm 1mm 3mm 1mm, clip]{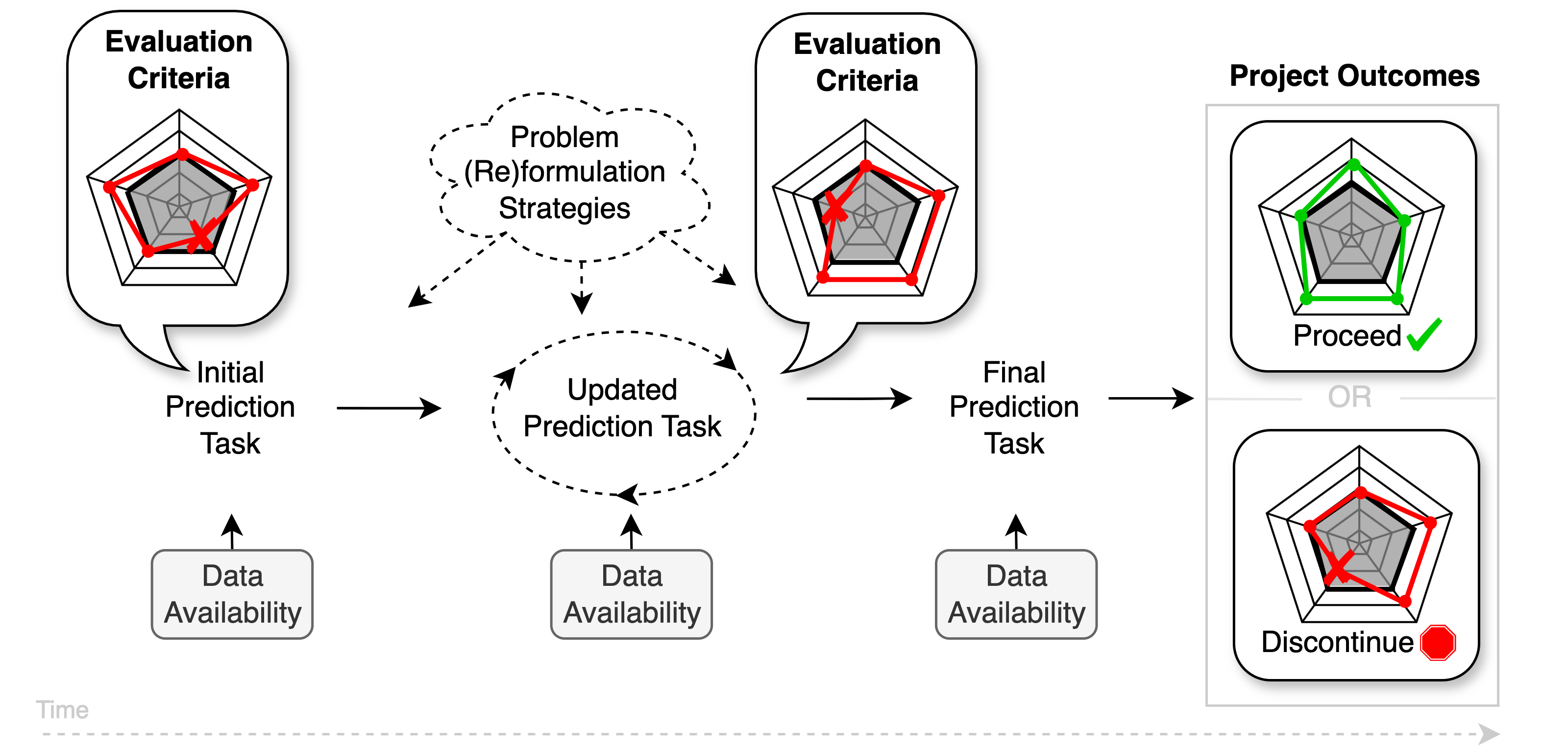}
    \caption{An illustration of the target variable construction process presented in our findings. During target variable construction, data scientists specify an initial prediction task based on their available data, then iteratively refine their prediction task by applying (re)formulation strategies. Data scientists proceed with their final prediction task if it satisfies all criteria, or discontinue their project if strategies are exhausted.\looseness=-1}
    \label{fig:main_findings}
\end{figure}

\newpage
\section{Findings}\label{sec:findings}

Our findings illustrate that target variable construction is a bricolage practice, in which data scientists design measurement instruments (i.e., target variables) for unobservable constructs by creatively making do with their existing data. While measurement has traditionally been conceptualized as a \textit{monologue} \citep{turkle2011life} --- a one-way, top-down translation from theoretical constructs to measurements in data --- we find that data scientists instead construct target variables through a rich process of \textit{dialogue} with their data, driven by a need to make do with resource constraints.\looseness=-1

The specific process of bricolage unfolds as  depicted in Figure \ref{fig:main_findings}. Data scientists begin by formulating an initial prediction task based on available data. They then iteratively inspect their formulation along several dimensions: ``Can the target variable be predicted with `reasonable' accuracy?'', ``Will the necessary data be available across deployment contexts?'' When issues arise along one or more criteria, data scientists apply various (re)formulation strategies to address these gaps. These strategies serve as the ``brushstrokes'' of data science work. Like a painter with a canvas, data scientists apply a strategy, step back to survey their progress along multiple criteria, then consider a new strategy in response to revealed flaws. The resulting problem formulation, while rarely ideal, reflects what the bricoleur considers ``fit for purpose'', by acknowledging imperfection while identifying creative and practical solutions that might never emerge through rigid top-down design. Data scientists eventually proceed with a formulation if they believe it satisfies their criteria across multiple dimensions, or discontinue their project if all strategies have been exhausted.\looseness=-1

In the following subsections, we detail this bricolage process by describing how data scientists balance validity with other competing criteria ($\S$ \ref{subsec:tradeoffs}), apply (re)formulation strategies ($\S$ \ref{subsec:strategies}), and the engage with theory-driven and data-driven evaluation activities to evaluate validity ($\S$ \ref{subsec:evaluating_construct_validity}).\looseness=-1

\begin{figure}[!t]
    \centering
    \begin{minipage}[t]{.8\textwidth}
        \centering
        \includegraphics[trim={5mm 6mm 5mm 5mm},clip,width=.7\textwidth]{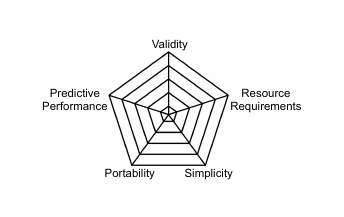}
        \caption{The five criteria data scientists evaluated during target variable construction.}
        \label{fig:eval_criteria}
    \end{minipage}
    \par\vspace{7mm}
    \begin{minipage}[t]{\textwidth}  
        \renewcommand{\arraystretch}{1.2}  
        \centering
        \begin{tabular}{>{\centering\arraybackslash} m{2.2cm} >{\raggedright\arraybackslash} m{4.4cm} >{\raggedright\arraybackslash} m{5.6cm}}  
           \toprule 
           \textbf{Criterion} & \textbf{Description} & \textbf{Illustrative Tradeoff}  \\ 
           \midrule
           \textbf{Validity} & The extent to which a model assesses the specific concept or construct that the data scientist intends to measure \citep{coston2023validity, drost2011validity}. & N/A \\
           \textbf{Predictive Performance} & The extent to which a model's predictions align with observed outcomes (e.g., accuracy or AU-ROC). & Predicting cardiovascular rather than renal diabetes health outcomes (P10). See $\S \ref{subsec:swapping}$. \\
           \textbf{\makecell{Resource \\ Requirements}} & The organizational resources required to implement and maintain the formulation. & Making do with limited test scores to avoid imposing additional testing (P6). See $\S$ \ref{ref:balancing_validity_resources}. \\
           \textbf{Portability} & The ease by which a task definition can be transferred across geographic or institutional contexts. & Using coarse veterinary health information to support model inference across clinical contexts (P13). \\
           \textbf{Simplicity} & The complexity of a task definition or model's structure and interpretation. & Switching from a complex multi-outcome model due to challenges in explanation (P2). See $\S$ \ref{ref:composition}. \\
           \bottomrule
        \end{tabular}
        \captionof{table}{A description of criteria that data scientists considered while evaluating their problem formulation. Right column provides an illustrative example of how participants balanced each criterion with validity.}
        \label{tab:eval_criteria}
    \end{minipage}
\end{figure}

\subsection{Balancing Validity With Other Criteria}\label{subsec:tradeoffs}

While engaging in target variable construction, data scientists were forced to make do with the data they already had available for the task at hand. This process of making do forced data scientists to navigate a rich space of tradeoffs between validity and other important criteria, such as predictive performance, simplicity, resource requirements, and the ease by which it could be translated to different geographic or institutional contexts (i.e., portability) (Figure \ref{fig:eval_criteria}). In Table \ref{tab:eval_criteria}, we provide an overview of illustrative tradeoffs that data scientists made between validity and other criteria. We provide a description of \textit{how} this process of making tradeoffs unfolded in the sections below.

\subsubsection{Balancing Validity with Simplicity} 

While constructing prediction tasks, several data scientists navigated a tension between the \textit{validity} and \textit{simplicity} of their formulation. In particular, data scientists found complex formulations, such as those combining multiple outcome variables, hard to explain to other stakeholders and maintain in production. For example, P1 was a director of a data science team at a large education company. Because the essay scoring tool designed by  their team was used by schools throughout the country, P1 described the \textit{simplicity} of  their formulation as critical for ensuring the system was robust and maintainable.  P1 \textit{had} considered adopting a sophisticated multi-target formulation to improve validity. Yet,  their team later abandoned the approach over practical concerns, explaining that the sophisticated models  \textit{``are a real hassle in practice. The extra [...] bells and whistles [...] made it flakier code, something that we didn't necessarily want to deploy at scale, compared to a relatively straightforward single output variable model.''} Further, P1 later shared that they felt it was possible to proceed with using a simple single output variable model if they made several hyperparameter adjustments and user interface tweaks. In this experience, P1 was willing to settle for a problem formulation that had minor validity drawbacks because they felt it would be simple and easy to maintain in a production system.

\subsubsection{Balancing Validity with Resource Requirements}\label{ref:balancing_validity_resources} 

Data scientists also evaluated the validity of their formulation against its resource requirements. For example, P6 was a data scientist developing an online math tutoring platform. P6 explained that the time and cost involved with acquiring ``gold standard'' student test scores to evaluate  their models was a frequent challenge. School districts often had data sharing restrictions. Opportunities to test students were limited. However, P6 explained that even if they \textit{could} test students more often, they would not recommend that approach. Testing students is \textit{``just removing opportunity. Removing time from the students' school calendar, which is already [...] crowded.’’} Instead of collecting data, P6 devised a creative workaround that achieved their project's goals using the data they \textit{already} had available.\footnote{We detail the strategy P6 and their team used in Section \ref{subsec:bridging}.} While P6 acknowledged that their workaround was imperfect, they were fine with proceeding so long as they had \textit{``at least some confidence''} in its validity. In particular, while reflecting on potential drawbacks of their modeling decisions, P6 shared that \textit{``this [application] isn't high stakes''} when compared to other potential modeling contexts (e.g., university admissions). They explained that had the model been deployed in other contexts, they might apply a different level of stringency to the validity of their approach.

\subsubsection{Elasticity of Evaluation Criteria}

Data scientists applied varying levels of flexibility to different evaluation criteria, a characteristic we term \textit{elasticity}. Participants frequently demonstrated elasticity with validity standards, tolerating imperfections when doing so reduced the complexity or resource requirements of their formulation. Data scientists \textit{calibrated} this tolerance based on the success of compensatory adjustments (e.g., user interface tweaks) and the perceived ``stakes'' of their modeling decisions. Interestingly, participants maintained more rigid standards for other criteria. For instance, several participants cited hard cutoffs for acceptable predictive performance --- e.g., P8 applied an AU-ROC threshold of 0.7 to determine whether to proceed with a project. This rigidity in predictive performance evaluation likely stems from the ease with which it can be distilled into a single, quantifiable metric, unlike validity, which requires a more holistic assessment involving multiple sources of evidence \citep{jacobs2021measurement}. While tradeoffs are an inherent component of bricolage, motivated by adaptive response to resource constraints, the process of \textit{navigating} tradeoffs remains poorly supported. We describe opportunities to help data scientists navigate tradeoffs in Section \ref{subsection:supporting_validity_tradeoffs}.

\subsection{Strategies for (Re)formulating Prediction Tasks}\label{subsec:strategies}

Given their limited and imperfect data resources, how do data scientists engage in bricolage to craft a formulation that satisfies their criteria? Much like a painter refining a painting through successive brushstrokes, or a programmer compiling and re-compiling a system after adjusting lines of code \citep{turkle2011life}, data scientists iteratively construct target variables through \textit{(re)formulation strategies}. These strategies form the backbone of the dialogue between the data scientist and their formulation, enabling them to respond adaptively to constraints encountered while working with existing data. At times, data scientists apply these strategies instinctively, without a second thought. At other times, data scientists encountered eureka moments after discovering a strategy that leveraged unexpected data sources or combined existing ones in novel ways. In the following sections, we detail five key strategies—piggybacking, composing, swapping, bridging, and refining—that data scientists used to navigate the constraints of their limited repertoire.

\subsubsection{Piggybacking}

During the initial stages of problem formulation, data scientists often constructed target variables through \textbf{piggybacking:} \textit{adopting or extending a problem formulation that was previously used in a similar context in the past}. Participants justified piggybacking by citing precedents from related projects. For example, while developing a model to inform university admissions decisions, P5 suggested predicting whether students' college GPA exceeded a 3.0 because other data science teams used it for similar admissions pipelines in the past. Piggybacking provided a rationale for P5's formulation decisions and offered a point of comparison with prior work. However, they were also wary that piggybacking could introduce unintended validity consequences.

One participant encountered a validity flaw in their formulation after piggybacking. P1 was developing an automated essay scoring tool on a six-point scale, where six is the best. P1 initially piggybacked by adopting the most prominent evaluation metric in the literature (Quadratic Weighted Kappa) but quickly realized it led to jarring results. 

\begin{quote}
    \textit{``The model simply never gave that top score. The more accurate [model] predicted the two’s, three’s, four’s, five’s. There are students out there that simply don't accept that. They'll write 30, 40, 50, 60 attempts at an essay, trying to get [...] a perfect score.''}
\end{quote}

Piggybacking off of other modeling contexts did not work for P1 - optimizing for Quadratic Weighted Kappa yielded a model that almost never gave a perfect score. This led P1 to eventually adopt a performance metric better suited to their setting, which balanced predictive accuracy with some chance that students could get a perfect score. P1's experience illustrates that top-down, theoretical analysis of an established precedent (Quadratic Weighted Kappa) was insufficient for making a well-informed piggybacking decision. Instead, P1's experience illustrates the ``science of the concrete'' characteristic of bricolage, where tangible interactions with materials are important for understanding how to assemble elements in the repertoire into a functional artifact.\looseness=-1

\subsubsection{Composing}\label{ref:composition}

Several data scientists constructed prediction tasks by \textbf{composing}: \textit{assembling multiple outcome variables into a single prediction target}. Often, data scientists used composing after realizing that a single outcome variable did not capture all dimensions of a construct they were trying to measure. For example, P8 used composing to predict the ``academic work ethic'' of college freshmen from their high school performance. While high school GPA was an established predictor of collegiate success, P8 felt that GPA alone would not offer enough specificity to differentiate work ethic from other success factors. P8 reasoned about how they might \textit{compose} standardized test scores and GPA in a single measure for ``academic work ethic'' by combining the two:

\begin{quote}
\textit{``If you assume that SAT or ACT scores are [...] indicators of academic knowledge, [...] this is where the person's at. If there's a difference between where the person should be and where they're at, we thought it was reasonable to conceptualize that as work ethic.''} 
\end{quote}

P8 designed a regression model that predicted GPA given test scores, creating a model that accounted for the gap between them to capture ``work ethic.'' A negative residual indicated that students had lower grades than predicted by their test scores, thereby signaling work ethic concerns. While teaching an introductory freshman class, P8 was surprised to find that students predicted to have poor work ethic were more willing than others to copy homework. Their model's predictions also closely aligned with students' long-term persistence outcomes (e.g., retention and graduation). These observations provide evidence for the \textit{consequential validity} of the model.

Yet some participants found that \textit{composing} had simplicity limitations. For example, P2 tried to use a multi-outcome latent variable model to predict student ``engagement.'' However, when P2 presented the model to software engineers and product managers, they distrusted its predictions when the model contradicted their prior beliefs about new product features. P2 explained that:

\begin{quote}
\textit{``If I show them a latent [variable] model, and it says, ‘No, it doesn't look like you improved engagement’ then there's gonna be a lot of questions about that model. What is this actually measuring? How do you explain it?''}
\end{quote}

P2 found it easier to justify their model when it targeted a single outcome based on what software engineers and product managers viewed as ``hard data.'' As a result, while P2 acknowledged validity benefits with latent variable models, they did not use them in practice. This tension between validity and simplicity highlights how bricolage solutions often bear the mark of compromise, as data scientists negotiate between theoretical ideals and practical constraints during problem formulation.\looseness=-1

\subsubsection{Swapping}\label{subsec:swapping}

As data scientists worked with their formulation, they sometimes reassessed their initial choice of outcome variable. The \textbf{swapping} strategy describes \textit{changing the outcome variable when the initial one fails to meet evaluation criteria}. For example, P10 used swapping while developing a model to predict diabetes-related ``health complications.'' P10 initially operationalized ``health complications'' using a cardiovascular outcome. Yet after discovering that cardiovascular outcomes were hard to predict, they later swapped to a measure of kidney function:

\begin{quote}
\textit{``At some point, after […] seeing that this [cardiovascular] condition is quite hard to predict, I thought that switching to something that is strongly related to the biomarkers I had available would make more sense. […] Seeing the data, getting to work with data, you might need to change your outcomes.’’}
\end{quote}

However, P10's use of renal outcomes was controversial because the renal biomarkers used as features were closely related to the outcome definition. As a result, it was not clear that the new model captured a true predictive signal in their data. This concern relates to \textit{internal validity}, or ``the existence of a defensible causal relationship between features and the target label'' \citep{coston2023validity}. However, the goal of this project was to demonstrate a new prediction method, so P10 tolerated internal validity limitations. P10 explained that they would place greater weight on these internal validity concerns if they were deploying their model in a clinical context. P10's experience embodies the responsive dialogue central to bricolage, where ``seeing the data, getting to work with data'' led to a reconsideration of the initial approach based on the artifact that could be constructed with the materials at hand.\looseness=-1

\subsubsection{Bridging}\label{subsec:bridging}

While working with their data, two participants identified ``gold standard'' outcomes with established validity that were rarely available and infeasible to directly predict. Data scientists addressed this challenge by \textbf{bridging}: \textit{targeting a low-cost, readily available outcome as a surrogate for the more expensive gold standard measure}. While bridging, participants evaluated their surrogate model using the gold standard outcome to construct a ``bridge'' between the surrogate model and the unobservable construct of interest.

For example, P6 used bridging to evaluate the ``readability improvement'' of content changes they made to an online learning platform. While P6 wanted to assess whether their changes helped struggling readers, they only had access to ``gold standard'' test scores from a small fraction of schools. P6 circumvented this issue by using students' performance on early lessons in the software as a surrogate for their reading ability:

\begin{quote}
    \textit{``We had this idea that came from another paper that […] this early lesson [...] was a good proxy for reading ability. So we built some [...] models that took raw interaction data and [we] found a neural network model that does a reasonable job of predicting student's [test] performance from this first lesson.''}
\end{quote}

P6 validated their surrogate model by checking its performance against the test scores they \textit{did} have available. Once they were confident that the model had ``reasonable'' validity, they used it to evaluate whether content changes helped students predicted to be poor readers. They found that the content updates offered a large readability improvement poor readers, declaring the project ``a really big success.'' P6 acknowledged that both the gold standard tests and the formulation were imperfect. However, given the analytics focus of the project, P6 had some tolerance for validity imperfections in their bridging formulation. This tolerance for imperfection, balanced against practical gains, exemplifies the bricoleur's pragmatic approach to problem-solving under resource constraints.\looseness=-1

\subsubsection{Refining}\label{subsec:refinement} During target variable construction, participants also tweaked their formulation by making small changes. This strategy involved \textbf{refining}: \textit{modifying the target variable definition in response to a defect along one or more criteria}. In contrast to the larger change in outcomes involved with swapping, participants used refining to make more granular formulation adjustments.

For example, P13 used \textit{refining} to narrow in on definitions for animal diseases in their dataset. While human healthcare had a mature ICD-9 disease classification scheme, P13 explained that animal health conditions remained vague and poorly specified. Clinicians were also unsure of how to map their understanding of pet diseases into a \textbf{label definition}: or a \textit{systematic set of classification rules in data}. P13 addressed this issue by refining the label definition with clinicians.

\begin{quote}
\textit{``Clinicians […] can't think of how do I define the dog has diabetes? I would have the vet look at it and [ask if] there [are] symptoms and signs [...] I couldn't get right. That takes many times of iteration to find out.’’}
\end{quote}

At each stage of refinement, P13 and the vet used written clinical notes about the pet to identify animals who were labeled incorrectly under the current definition. P13 then revised the definition to correctly categorize the misclassified instances (e.g., by adding a new rule to the definition based white blood cell count). These iterations continued 20 to 30 times until the label definition sufficiently captured the disease. P13 emphasized a need for practicality during the process: they were fine with 10\% of diseases being mislabeled if the major dimensions of the disease definition were covered. P13's acceptance of imperfection while ensuring ``major dimensions'' were covered typifies the bricoleur's focus on creating solutions that are ``fit for purpose'' rather than ideal.

\subsubsection{When Reformulation Fails}

After applying strategies, many data scientists eventually crafted a prediction task that satisfied their criteria. However, one discontinued a project after identifying a flaw that they could not patch through reformulation. P12 was developing a model to flag patients in the general care clinic who were at risk of Celiac disease. An initial model predicted test results with good accuracy. However, after carefully inspecting their formulation, P12 realized that patients with test results already appeared sick to physicians. Training a model on patients who \textit{appeared sick} then deploying the model to \textit{the full patient population} threatened the model's external validity. P12's team considered a variety of workarounds to patch this gap. However, when no solutions appeared feasible, P12 discontinued the project. While most data scientists applied strategies in response to visible defects in their formulation (e.g., predictive performance concerns), the flaw identified by P12 was \textit{invisible}. In particular, identifying this issue required scrutinizing the connection between the dataset available for model development (i.e., patients who \textit{appeared} sick to doctors) and the population at deployment time (i.e., the general population entering the clinic). We describe other approaches that data scientists used to shed light on invisible validity defects in Section \ref{subsec:evaluating_construct_validity}.

\subsection{Evaluating the Validity of a Problem Formulation}\label{subsec:evaluating_construct_validity}

While applying strategies, data scientists often took a step back to evaluate the validity of their problem formulation. Drawing on domain knowledge, they engaged in theory-driven practices to probe conceptual relationships between their target variable and theoretical constructs of interest. However, data scientists also treated their problem formulation as a material object to be  scrutinized by inspecting concrete predictions generated on specific data instances. Lévi-Strauss calls this more tangible approach ``the science of the concrete'' --- whereby the bricoleur understands their artifact through direct manipulation of material elements as opposed to abstract, conceptual reasoning. We now describe these theory-driven and data-driven practices in detail, as summarized in Table \ref{tab:eval_approaches}.

\subsubsection{Theory-Driven Evaluation Practices}\label{subsec:spurious_eval}

While engaging in theory-driven practices, data scientists used their domain knowledge to reason about a problem formulation \textit{conceptually}, without scrutinizing a specific data sample. Theory-driven practices were informed by participants' prior modeling experience and an understanding of the sociotechnical context (e.g., hospitals, classrooms) in which their models operate. Though participants had varying levels of formal measurement theory training, many theory-driven practices connect to established construct validity sub-criteria. 

While engaging with the problem formulation vignette, several participants identified spurious factors that might affect an outcome variable while remaining unrelated to the construct of interest. When considering ``length of hospital stay'' as a measure for ``health acuity,'' P11 noted that \textit{``there's so many other things that could contribute to somebody being in the emergency department for 12 hours that might not have anything to do with how sick they are''}, citing staffing levels and emergency department crowding as examples. In validity theory, this reasoning connects to \textit{divergent validity}, or the extent to which a problem formulation measures aspects of other unrelated constructs \citep{coston2023validity}.

Several participants used multiple outcomes to develop a more complete understanding of how a construct should be conceptualized. For example, P2 wanted to understand whether students who used hints were ``gaming'' the lesson -- i.e., using hints to finish questions without engaging with the material. Accuracy did not provide enough information to identify gaming from hint use because students could answer a question incorrectly, even after using a hint and applying effort. However, P2 found that students took longer to respond when they were engaged with the lesson. As a result, a combination of \textit{accuracy}, \textit{response time}, and \textit{hint use} was necessary to triangulate gaming behavior. Reasoning across multiple outcomes connects to an assessment of content validity, or the extent to which an operationalization captures the substantive nature of the construct \citep{jacobs2021measurement}.\looseness=-1

\begin{table*}
  \renewcommand{\arraystretch}{1.1} 
  \begin{tabular}{>{\centering\arraybackslash} m{1.6cm} >{\centering\arraybackslash} m{2.2cm} >{\centering\arraybackslash} m{3.4cm} >{\centering\arraybackslash} m{5.3cm} } 
     \toprule 
     \textbf{Type} & \textbf{Practice} & \textbf{Description} & \textbf{Example}  \\ 
     \midrule
      & \textbf{Identifying spurious causes of outcomes.}  & Identifying factors influencing the outcome unrelated to the construct of interest.  & \textit{``There's so many other things that could contribute to somebody being in the emergency department for 12 hours that might not have anything to do with how sick they are.'' (P11)}  \\ 
      
      Theory-Driven & \textbf{Reasoning across multiple outcomes.}   & Leveraging multiple outcomes to gain a more holistic understanding of a construct. & \textit{``If you have high response times, and they're using [learning] supports, that's a good indication that they're trying, even if they don't get the next problem correct.'' (P2)} \\ 
      
      & \textbf{Linking interventions with outcomes.}  & Reasoning about potential causal relationships between an intervention and outcomes.  & \textit{``To get in the neighborhood of the right outcome measure, it is often like, well, what does this intervention really look like? […] Where would we ... expect to move the needle?'' (P6)} \\

      \hline

      & \textbf{Validating predictions in real-world deployments.}  & Checking the extent to which model predictions align with expectations in real-world conditions.  & \textit{``We did an evaluation study where the clinicians would actually review cases while the information that our models predicted would be relevant was highlighted for them.'' (P15)}  \\  

     Data-Driven & \textbf{Poking holes in label definitions.}   & Attempting to identify instances that are misclassified under a current label definition. & \textit{``[It was] a lot of literally, in some cases, printing off student essays and handing them to employees that were former teachers and saying, why did this person give this label?'' (P1)}  \\

      & \textbf{Applying quick checks and heuristics.}  & Using pragmatic checks to quickly gauge whether the task setup matches the data sample. &  

     \textit{``There's several models that existed in the literature that I attempted to transfer to my work. I checked it […] using finger in the wind sort of [approaches]. Like, what's the distribution of my data? Is it closer to 7\% or is it closer to 20\%?’’ (P7)} \\  
  
     \bottomrule
    \end{tabular}
    \caption{Data scientists engaged in both theory-driven and data-driven practices while evaluating validity.}\label{tab:eval_approaches}
    \clearpage
    \vspace{-2mm}
\end{table*}


Finally, participants described their choice of outcome variable as closely linked to the intervention being informed by a model. For example, P6 needed to decide which outcome time horizon (i.e., question lesson, week, semester, or year) was most appropriate for measuring the affect of platform content changes on student learning. P6 shared that \textit{``to get in the right neighborhood, where would we reasonably expect to move the needle?''} If the intervention included many content changes, they leaned towards an end-of-year student performance outcome. However, if a change to the platform was more granular, they would start at a more fine grained question or lesson level. This assessment of the relationship between interventions and outcomes connects to \textit{consequential validity}, or ``the real-world consequences of using the measurements obtained from a measurement model'' \citep{jacobs2021measurement}.\looseness=-1

\subsubsection{Data-Driven Evaluation Practices}

While engaging in data-driven practices, participants checked whether their theoretical understanding of a construct was borne out by the measurements they observed on specific data instances. Gaps between theoretical expectations and observations signaled a potential flaw in their formulation to be scrutinized. By identifying, probing, and later reconciling these gaps through a strategy, data scientists slowly built trust in their formulation's validity. In some cases, participants checked how a target variable categorized specific data instances (e.g., ``ground truth'' labels constructed by applying a label definition). In others, they checked the downstream predictions generated by a model trained on a target variable. Each of these practices depended heavily upon the specific data sample used to generate measurements. As such, our understanding of these activities emerged through the directed storytelling segment of interviews, as participants recounted specific times they manipulated a dataset in prior projects.

Several participants applied \textit{heuristics} to test whether measurements were inline with expected theoretical properties. For example, while evaluating new test items, P6 plotted students' scores as a function of their question attempt. A \textit{``nice, monotonically increasing curve''} with a \textit{``logarithmic flavor''} provided evidence that the item encoded knowledge of a single skill. P9 checked that the base rate of a disease derived from diagnostic codes matched the expected prevalence. A large mismatch was a signal that piggybacking off of existing disease definitions was infeasible. Participants described these heuristics as imperfect; a necessary but insufficient condition for demonstrating validity. These checks and heuristics are an example of evaluating \textit{face validity}, or the degree to which an operationalization measures what it purports to measure based on a quick, surficial assessment \citep{coston2023validity}.\looseness=-1

Participants tried to \textit{poke holes in label definitions} by identifying cases a label definition miscategorized the instance in comparison to theoretical expectations. While \textit{refining}, P13 tried to poke holes by identifying specific cases where the expert judgment of the clinician diverged from the categorization assigned by the label definition.
In education, P1 tried to poke holes in essay scoring rubrics by printing off physical copies and asking employees who were former teachers to explain \textit{``why did this person give this label?''} Employees' response served to build P1's intuitive sense of \textit{how} the rubric was used to assign instance-level scores, and identify potential gaps. By \textit{``staying close to the data''}, P1 had the trust needed in the labeling process to justify specific essay scores to students, teachers, other external stakeholders. P1 described this trust as critical should a measurement decision later require formal legal justification --- e.g., in the context of university admissions. For both P1 and P13, poking holes required working in tandem with domain experts, who had the knowledge required to verify whether the instance \textit{really was} positive or negative for the construct of interest. The practice of poking holes and patching them connects to an assessment of \textit{content validity}.\looseness=-1 

Finally, participants checked whether model behavior reflected their expectations in real-world deployment contexts. P15 checked the validity of a model trained to predict the ``clinical relevance'' of electronic medical records by conducting a field study that tested whether predictions saved clinicians time. Time savings provided evidence that the predicted fields really were ``clinically relevant.'' P15 and many other participants found both theory-driven and data-driven evaluation practices critical for establishing a holistic understanding of their formulation's validity.

\section{Discussion}\label{sec:discussion}

Today, significant focus across the HCI, CSCW, ML, and FAccT communities has turned towards evaluating the extent to which algorithmic systems achieve their purported aims. Given this attention, researchers and practitioners increasingly draw upon validity concepts established in the quantitative social sciences to describe desirable properties of algorithmic systems. For instance, a growing body of work examines how validity sub-criteria --- such as content, convergent, and discriminant validity --- might be applied to evaluate algorithmic systems \citep{coston2023validity, kawakami2024situate, guerdan2023ground, jacobs2021measurement, raji2022fallacy, liu2024ecbd, raji2021ai, alaa2025medical}. Yet our findings uncover a disconnect between the traditional, top-down measurement process assumed by validity theory and the ways data scientists navigate measurement decisions in practice. In particular, data scientists typically must make do with limited resources by re-appropriating existing data for new measurement tasks. These constraints force data scientists to balance validity with other important criteria, such as predictive performance, simplicity, and resource requirements. 

Based on these findings, we argue that \textbf{the challenge is not to \textit{replace} these bricolage practices with top-down planning, but rather to develop forms of scaffolding that can enhance the rigor of bricolage while preserving its creativity and adaptability.} Our findings provide a foundation for the research community to develop this scaffolding, by identifying concrete opportunities to help data scientists balance validity with other criteria, identify and apply problem (re)formulation strategies, and assess validity using both top-down and bottom-up  evaluation practices. We now reflect on data science as a bricolage practice, before providing targeted recommendations for supporting data scientists' problem formulation practices going forward.\looseness=-1

\subsection{Data Science as Bricolage}

References to data science often invoke engineering metaphors: data scientists ``engineer'' features, ``build'' models, and ``architect'' computational solutions. Yet our findings suggest that target variable construction more closely resembles bricolage than engineering—a mode of knowledge production whereby the bricoleur makes do with the resources at hand rather than acquiring materials for a predefined top-down plan. Our findings highlight how this process of making do unfolds in real-world projects, as data scientists re-appropriate data fields originally collected for entirely different purposes to accomplish their measurement goals. Bricolage describes \textit{how} data scientists orient themselves towards measurement problems, as well as nature and quality of their solutions.

\subsubsection{The Data Science Repertoire}

 The bricoleur operates with an inventory of odds and ends called a ``repertoire.'' Data scientists' repertoire includes both material resources (e.g., datasets, computational tools) and immaterial resources (e.g., domain theories, methods) brought to bear on a modeling task. The repertoire shapes the space of possibilities that bricoleurs imagine while confronting new problems. As \citet{louridas1999design} observes while drawing connections between bricolage and design, ``the bricoleur asks his collection, whereas the engineer, like the scientist, asks the universe.'' P12 eloquently translated this to data science, lamenting that \textit{``ultimately, you can't create data that don't exist.''} The constrained nature of data scientists' repertoire forces them to re-examine their limited materials from new perspectives to identify new and innovative solutions.\looseness=-1

Materials in data scientists' repertoire function as what Lévi-Strauss calls ``signs'', or concrete objects that represent abstract concepts, but remain constrained by their imagined uses. P6's creative re-purposing of raw user interactions in an online learning platform as a proxy for ``reading ability'' exemplifies the use of a sign, by unexpectedly transforming a mundane data field into an operationalization of a construct. Yet, as \citet{lvi1966savage} notes, ``the possibilities remain always limited by the particular history of each piece, and by what is predetermined in it due to the original usage for which it was conceived.'' Had P6 \textit{not} been aware of this proxy for ``reading ability'', established through a chance encounter with a colleague's work, they may have never made this connection.\looseness=-1

\subsubsection{Engaging in Dialogue with Data}

After taking stock of the repertoire, the bricoleur engages in dialogue with their materials to identify a workable solution to the problem at hand. As Lévi-Strauss explains, the bricoleur must ``engage into a kind of dialog with [the repertoire], to index, before choosing among them, the possible answers that the set can offer to his problem.'' \citep{lvi1966savage} This dialogue not only shapes the bricoleur's understanding of each element in isolation, but also their interactions: ``one element's possibilities interact with all other elements' possibilities, with the overall organization of the artifact he makes. The results of these interactions are never what he expects, and he must respond to them.'' \citep{louridas1999design} P10's experience illustrates how dialogue unfolds in data science. After initially selecting a cardiovascular outcome based on precedent in academic literature (piggybacking), \textit{``seeing the data, getting to work with data''} led them to \textit{swap} to a renal outcome that was more predictable given available biomarkers. This example illustrates problem (re)formulation strategies serve as the brushstroke of measurement in data science work.  Like a painter with a canvas, data scientists apply a strategy, step back to survey their progress along multiple criteria, then consider a new strategy in response to newly-identified flaws.

\textit{Dialogue} marks a notable shift from the top-down planning approach to measurement. Under top-down planning, a practice better described as a \textit{monologue} \citep{turkle2011life}, the data scientist defines a theoretical construct of interest, designs an operationalization, \textit{then} collects data. Yet this monologue supposes that the data scientist can navigate the space of constraints and contingencies in their formulation through mental simulation, without a specific data sample or model. Our findings underscore that this is rarely the case. Instead, data scientists required \textit{tangible interactions} with data to understand its affordances. It was through this very process of tangible interaction with data, borne out of resource constraints, that data scientists made creative leaps in their projects. As P6 explained, had a ``magical oracle'' granted their team access to all the data they required, they would have simply collected a comprehensive dataset with test outcomes. It was the combination of resource constraints and tangible interaction with data that drove P6 to \textit{bridge} using the limited data at hand.\looseness=-1

\subsubsection{The Emerging Artifact}
Critically, the artifact resulting from bricolage is rarely ideal, but rather bears the marks of the constraints, compromises, and creator that shaped it. As \citet{louridas1999design} explains, ``bricolage is ... at the mercy of contingencies, either external, in the form of influences, constraints, and adversities of the external world, or internal, in the form of the creator's idiosyncrasy.'' Analogously, data scientists  readily acknowledged imperfection in their solutions. P13 explained that a label definition that miscategorized 10\% of pet diseases was acceptable if the ``major dimensions'' of the disease's theoretical conceptualization were covered. P6 deemed their re-purposing of existing data a ``really big success'', but also identified important limitations. Thus, from the bricoleur's perspective, the key question is not whether a solution is \textit{ideal}, but rather whether it is \textit{fit for purpose}. P6 embodied this ethos when they justified \textit{bridging} by explaining that their platform analytics use case was ``not high stakes.'' Had their model been designed for use cases they perceived as higher-stakes, P6 explained that they may have applied a higher level of stringency.\looseness=-1

Beyond constraints, Lévi-Strauss notes a personal dimension to the artifact that emerges from bricolage: ``he [the bricoleur] `speaks' [...] through the choices he makes among the limited possibilities, the character and the life of the creator.'' P8's creative use of \textit{composing}, which conceptualizes ``academic work ethic'' as the difference between \textit{where a student is at} and \textit{where they should be}, may have never emerged had they not also observed real students in real classrooms as an educator. Similarly, P12's project, which had strong buy-in from multiple stakeholders and momentum towards completion, never came to fruition due to a validity flaw spotted by P12. Yet it was through P12's personal interest in adjacent bodies of biostatistics literature, and their attention to methodological rigor in problem formulation, that they became aware of this otherwise invisible flaw. As these experiences illustrate, the final problem formulation depends greatly on the experience of the bricoleur.\looseness=-1 

In sum, bricolage acknowledges that data scientists must work within the confines of their existing resources. This practical consideration is neither centered nor prioritized in the conventional top-down approach to measurement. Thus, short of dramatically expanding the available resources for a project, a rigid enforcement of top-down planning is unlikely to translate to a desirable project outcome. \textbf{Therefore, rather than recommending more stringent enforcement of the traditional top-down measurement approach, we suggest more effectively scaffolding \textit{existing} bricolage practices.} We now revisit traditional top-down measurement interventions with this insight in mind, before envisioning new ways of scaffolding data science practice.\looseness=-1

\subsection{Design Implications: Reconciling Top-Down Measurement Interventions with Bricolage Practice}

Our findings uncover a tension between existing interventions designed to help analysts implement the traditional top-down measurement workflow, and data scientists' bottom-up bricolage practices. However, this tension is not unique to data science. History reveals similar patterns across domains where top-down planning approaches have failed while bottom-up alternatives succeeded. In urban planning, modernist urban renewal projects that imposed regimented public works projects onto complex social systems often destroyed functioning neighborhoods, while community-based planning approaches showed strong success \citep{fuller2017analysis, scott2020seeing}. In software development, rigid and systematic waterfall planning methodologies gave way to agile approaches that embrace iteration and adaptation \citep{osorio2011moving}. Key to the success of these bottom-up alternatives was not wholesale rejection of structure, but rather targeted interventions like mixed-use zoning in urban planning and sprint retrospectives in agile development. These interventions recognized that while existing practices were in need of reform, effective solutions must work within resource constraints and pinpoint specific procedural weaknesses in need of constructive redirection. Similarly, when improving measurement practices in data science, interventions tailored to support data scientists' \textit{existing} practices are likely to be more effective than direct translation of rigid, top-down measurement frameworks. \lgdelete{}\lgedit{In the following sub-sections, we examine how existing top-down measurement interventions might be adapted to better support data scientists' bricolage practices.}

\subsubsection{Pre-Registration for Predictive Modeling.}

Pre-registration is one top-down intervention that is widely recommended in the quantitative social sciences \citep{van2016pre}. This protocol discourages data-dependent decision-making by requiring an analyst to specify outcome measures and hypotheses \textit{before} running analyses. Recent work argues for the adoption of pre-registration in predictive modeling contexts  \citep{hofman2023pre} --- i.e., by requiring data scientists to finalize their problem formulation before training a model. Yet requiring such pre-registration is akin to \textit{blindfolding the bricoleur} --- it restricts the very bottom-up, iterative process that data scientists require to make do under resource constraints. Indeed, data scientists interviewed by \citet{hofman2023pre} worried that pre-registration might limit their creativity, or make it challenging to later tweak their formulation in response to new challenges.\looseness=-1 

\lgedit{
A more targeted protocol for limiting data-driven decision-making might encourage teams to establish minimum performance standards for each criterion before beginning problem formulation. This approach could help to prevent \textit{standard erosion} – or a selective reduction in stringency applied to a criterion over subsequent rounds of (re)formulation. Our interviews surfaced evidence for standard erosion in several projects, as participants weakened validity standards to improve predictive performance or simplicity. To mitigate standard erosion, HCI and CSCW researchers might co-design tools such as radar plot visualizations (e.g., Figure \ref{fig:eval_criteria}) to help teams chart trade-off spaces across criteria and identify suitable operating regions. For instance, a research team might prioritize validity over simplicity, while an educational technology company might be more willing to accept validity trade-offs for improved simplicity. By making performance trade-offs explicit, teams can (1) prevent standard erosion, (2) enable iterative (re)formulation, and (3) provide rationale for discontinuing projects when minimum standards cannot be satisfied with available resources.
}

\subsubsection{Latent Variable Models}
Researchers in the quantitative social sciences have also developed \textit{statistical tools} to improve the validity of models. For example, latent variable models (e.g., \citep{bishop1998latent, bartholomew2011latent}) use advanced statistical machinery to characterize the relationship between unobservable theoretical constructs and observed proxies. Recent work has proposed latent variable models (e.g., \citep{milli2021optimizing}) and related modeling advances (e.g., \citep{watson2023multi, guerdan2023counterfactual}) to improve the validity of models used for decision support and recommendation systems. Yet these sophisticated approaches require advanced statistical training, multiple outcome variables, and significant implementation time. This imposes practical barriers to adoption. For example, P1 shared that, while they were aware of advanced multi-outcome modeling approaches, their ``extra bells and whistles' made them unreliable in production.\looseness=-1

This disconnect between theoretical benefits of latent variable models and practical constraints presents an opportunity to develop modeling tools that better align with bricolage practices. Rather than introducing increasingly complex modeling tools, future work might focus on \textit{simplifying} latent variable models to preserve their validity benefits while reducing implementation complexity. These simplified approaches could emphasize interpretability and robustness, helping teams effectively communicate their models to stakeholders without advanced statistical training. We elaborate further on improved modeling approaches designed to better facilitate bricolage in Section \ref{subsection:applying_strategies}.\lgdelete{, we elaborate on specific modeling approaches that are likely to be most effective for supporting bricolage practices, including simplified approaches to composing multiple outcomes that maintain validity while meeting simplicity constraints.}

\subsubsection{Pedagogical Interventions}\label{subsection:pedagogy}

Our findings also highlight opportunities for pedagogical interventions to better equip practitioners to navigate target variable construction. A key insight from our interviews is that data scientists often draw on prior experience, engaging in analogical reasoning by identifying parallels between their current project and those encountered previously. This highlights the value of \textbf{case-based learning} in data science education. Case-based approaches enable students to analyze concrete case studies and understand how experienced data scientists make trade-offs between criteria like validity, simplicity, and resource requirements. \lgedit{Such case-based approaches could build upon existing protocols designed to help practitioners identify potential failure modes of AI systems during the early stages of AI system design \citep{saxena2025ai, kuo2023understanding, jung2025making}}.

\textbf{Reflective practice} and \textbf{hands-on workshops} offer complementary roles in developing data scientists’ capacity to reason about validity. \lgdelete{Iterative experimentation and reflection help students develop a nuanced understanding of the consequences of methodological choices, enabling them to better justify and communicate those choices.} This mirrors how tasks like feature engineering are often described—as a “craft”—involving domain expertise, theoretical grounding, and direct engagement with data \citep{duboue2020art}. Workshops focused on (re)formulation strategies—such as piggybacking, composing, swapping, bridging, and refining—can expose students to the tactical, trial-and-error process of shaping a prediction task within real-world constraints. Together, these approaches offer an integrated method for teaching the complexities of bricolage-based problem formulation. 



\subsection{Design Implications: Developing a Scaffolding for Effective Bricolage Practice}\label{subsec:bricolage}

In addition to re-visiting existing top-down interventions, our findings surface a need for entirely \textit{new} interventions that center bricolage as a data science practice. While many toolkits help data scientists engage in inner-loop activities, such as training models \citep{pang2020deep, dingen2018regressionexplorer,cabrera2023zeno, yan2021tessera} and exploring data \citep{wongsuphasawat2015voyager, kery2018story}, data scientists lack tools to help them weigh validity with competing criteria, identify and apply strategies, and evaluate validity. \textbf{We now use our findings as a concrete scaffolding for identifying actionable points of intervention in data scientists' bricolage practices}. These interventions are intended to \textit{facilitate} key practices that data scientists found critical during problem formulation, while \textit{adding guardrails} around others that have the potential to introduce flaws.\looseness=-1

\subsubsection{Helping Data Scientists Weigh Tradeoffs}\label{subsection:supporting_validity_tradeoffs}

In the quantitative social sciences, validity is traditionally viewed as the primary criterion for the scientific integrity of a study. \lgedit{However, while engaging in bricolage, data scientists were forced to adaptively balance the validity of their formulation with competing criteria. Yet this balancing act remains poorly supported, leaving the (re)formulation processes susceptible to inadvertent introduction of validity defects.}

First, our findings illustrate a need to help data scientists design measures for their formulation's simplicity, validity, portability, and resource requirements. Teams currently rely heavily on measurements of predictive performance (e.g., AU-ROC or F1 scores) while applying strategies. Echoing the classic adage ``what gets measured matters'' \citep{skogan1999measuring}, this practice leads teams to prioritize predictive performance improvements at the expense of other criteria. To address this imbalance, HCI and CSCW researchers might work with teams to co-design rubrics that help them operationalize measures for specific validity sub-criteria (e.g., convergent, divergent, content), simplicity (e.g., model complexity, implementation requirements), and portability (e.g., data field availability across contexts). \lgedit{For instance, a team might use such a rubric to set acceptable inter-rater reliability thresholds between automated essay scoring and human expert ratings.} To support ease of implementation, this rubric could be paired with existing Responsible AI toolkits designed to help teams weigh the suitability of predictive modeling tools for specific applications \citep{kawakami2024situate, coston2023validity}.

Second, there is an opportunity to develop tools that help data scientists identify targeted data collection opportunities. While engaging in bricolage, data scientists were forced to weigh the predictive performance and validity benefits of data collection against its resource requirements. Yet existing model training and evaluation toolkits often assume that datasets are fixed \citep{cabrera2023zeno, yan2021tessera, bellamy2019ai}. Future HCI and CSCW work might close this gap by developing tools that help data scientists quantify the improvements offered by collecting additional data. \lgedit{The framework of data valuation offers tools for ascribing monetary value to data resources \citep{castro2023data, perdomo2023relative}. Future toolkits could build upon these frameworks to enable teams to estimate the benefits of new data sources prior to acquisition, potentially leveraging synthetic data generation to quickly assess potential improvements before collecting real data. These efforts would be enhanced by empirical studies investigating how teams acquire and ascribe worth to data throughout the (re)formulation process}.

\subsubsection{Helping Data Scientists Identify Strategies}\label{subsection:identifying_strategies}

Our findings present an opportunity to help teams \textit{identify} strategies \lgedit{during the target variable construction process}. Data scientists currently identify strategies in an ad hoc, unstructured manner \lgedit{-- e.g., by drawing upon} their bespoke knowledge of academic literature, or the expertise of others on their team. This approach led teams to miss opportunities to improve their formulation. For instance, following the conclusion of their project, P12 identified a modeling framework---i.e., case-control methods \citep{schlesselman1982case}---that was tailor-made to address their project's data constraints. Yet P12 was not aware of this framework because their background in machine learning did not involve training in biostatistics methods. This presents an opportunity to broaden data scientists' awareness of relevant strategies before helping them narrow in on which are best suited to their bespoke modeling challenges.

\lgedit{Future HCI and CSCW research might develop systems that help teams exchange knowledge surrounding common (re)formulation challenges. While community-based knowledge exchange is central to inner-loop workflows, as illustrated by Stack Overflow \citep{barua2014developers} for coding and Hugging Face \citep{shen2024hugginggpt} for model training, data scientists' support for problem formulation remains siloed. Similar to other online communities, this system might structure discussions via a comment forum where teams share specific strategies—P8's \textit{combining} strategy for measuring ``academic work ethic'' or P6's \textit{bridging} strategy for measuring ``reading ability.'' This forum could eventually be used to inform the design of consensus-backed protocols for navigating common (re)formulation challenges.}\looseness=-1

\subsubsection{Helping Data Scientists Apply Strategies}\label{subsection:applying_strategies}

\lgedit{Our findings also highlight opportunities to help data scientists \textit{apply} each strategy they identify more effectively. Therefore, we now describe opportunities to develop tools and processes that support data scientists in applying each strategy effectively, while also avoiding potential strategy-specific complications.}

\textbf{Piggybacking}. While \textit{piggybacking}, data scientists port an existing problem formulation to their own modeling context. Piggybacking saves time when there is a strong alignment between established precedent and data scientists' modeling context. Yet, as illustrated by P1's experiences, piggybacking can also introduce serious defects when contextual differences are overlooked. Going forward, there is a need for structured protocols that help teams thoughtfully reason about the transferability of problem formulations between modeling contexts. While many guidelines support dataset and model documentation (e.g., \citep{mitchell2019model, gebru2021datasheets, coston2023validity}), none support assessments of problem formulation transferability. HCI and CSCW researchers might fill this gap by working with data scientists to co-design a rubric that teams can complete while considering piggybacking. This rubric might guide data scientists through a side-by-side comparison of an existing formulation and their own along dimensions such as high-level modeling goals, available data, and rationale for a selected target variable. Helping data scientists spot transferability gaps can also help teams provide a rationale to external stakeholders should a team decide to use a formulation that deviates from an established precedent.\looseness=-1 

\textbf{Composing.} While \textit{composing}, data scientists combine multiple outcomes into a single target variable. Composing can improve validity by helping data scientists capture all relevant dimensions of a construct in a target variable. Yet composing can also introduce simplicity limitations. This tension indicates a need for simple and interpretable mechanisms for combining outcomes. Future HCI and ML research might explore opportunities for interpretable multi-outcome models that cleanly map conceptual relationships (e.g., \textit{``the difference between where the person should be and where they're at''}, P8) into transparent model designs. One approach likely to be effective involves training multiple models to target multiple outcomes of interest (e.g., cost of medical care, re-admission), then aggregating predictions into a single score at run-time. P11 described such an approach as common in the medical industry, where both simplicity and validity are paramount. Several participants also suggested this approach during the design probe. HCI, ML, and Visualization researchers can implement this approach by developing tools that help teams encode their domain knowledge into aggregation functions that weight multiple predictions into a single score.\looseness=-1

\textbf{Swapping.} While \textit{swapping}, data scientists switch to a different outcome after they identify defects in one or more criteria. Swapping can be adaptive if it helps teams identify a formulation that satisfies all minimum performance standards. Yet swapping can also introduce standard erosion. This tension highlights a need for tools to help data scientists make more systematic swapping decisions. While many existing systems support feature engineering (e.g., \citep{bhattacharya2024exmos, guo2011proactive, kandel2011wrangler, rattenbury2017principles}) and EDA (e.g., \citep{wongsuphasawat2015voyager, kery2018story}), only one supports \textit{outcome exploration} \citep{gala2024fairtargetsim}. HCI and Visualization researchers can support outcome exploration by developing toolkits that help data scientists weigh the portability, simplicity, predictive performance, and resource requirements of alternative outcomes. Because our results suggest that swapping can be enticing given its predictive performance benefits, it is especially important that toolkits help teams identify validity drawbacks of swapping decisions. We discuss opportunities to support validity evaluation while considering swapping in Section \ref{subsection:supporting_validity_evaluations}.

\textbf{Bridging}. While \textit{bridging}, data scientists adopt a low-cost, readily available target variable as a proxy for a more costly gold standard measure. Bridging exemplifies bricolage because it involves a creative approach for making do with limited data. Yet the success of bridging hinges on data scientists' ability to identify trustworthy sources of auxiliary information. To support bridging, data scientists need tools to help them identify, evaluate, and incorporate relevant sources of auxiliary data. While HCI and Data Engineering research has developed tools supporting data discovery (e.g., \cite{fernandez2018aurum, nargesian2019data, paton2023dataset, bogatu2020dataset}), existing toolkits are not tailored to support bridging. Future research might close this gap by developing tools that help data scientists forage for relevant auxiliary data and evaluate its expected utility against the resources required to incorporate it in their formulation. As with swapping, it is important that such tools help data scientists identify any defects associated with bridging.\looseness=-1

\textbf{Refining.} While \textit{refining}, data scientists make granular adjustments to how a target variable is coded in their data. Participants often engaged in refining by modifying a label definition used to operationalize a construct using a set of indicators used to categorize instances. For instance, P13 engaged in refining by working with domain experts to spot gaps in label definitions, then patch them with a revised set of indicators. Though refining can improve validity, practitioners lack tools to encode domain experts' knowledge into label definitions. Instead, existing tools developed by the HCI and ML communities help annotators apply a \textit{fixed} label definition to annotate data (e.g., \citep{doccano, muller2021designing, desmond2021increasing, daudert2020web}) or use crowdsourcing to refine annotation guidelines (e.g., \citep{chen2023judgment, bragg2018sprout, chang2017revolt, k2019taskmate, manam2018wingit}).\footnote{Traditional annotation guidelines are \textit{de facto} label definitions if they serve as a measurement instrument that operationalizes a latent construct of interest.} Future HCI and ML research can help data scientists encode domain experts' knowledge in label definitions by developing tools that help teams  collaboratively systematize concepts~\cite{adcock2001measurement} by iteratively identifying and refining indicators used in label definitions. Because our results indicate that domain experts require rich contextual information to identify areas where the systemization of a concept is incomplete, tools supporting this process should capture rich contextual information (e.g., written notes) in data.\looseness=-1

\subsubsection{Helping Data Scientists Evaluate Validity}\label{subsection:supporting_validity_evaluations}

\lgedit{

Disciplines in the quantitative social sciences have established numerous guidelines for evaluating the validity of a measurement instrument. Such guidelines are often grouped into validity sub-criteria --- such as convergent, divergent, and predictive validity --- describing abstract, theoretical assertions about the expected behavior of a measurement instrument \citep{drost2011validity}. Yet our findings illustrate a disconnect between these abstract assertions and the more concrete theory- and data-driven evaluation activities performed by practitioners. We resolve this tension by identifying opportunities to more effectively integrate measurement principles into data scientists' \textit{existing} evaluation practices (Table \ref{tab:eval_approaches}).
}



First, our findings highlight opportunities to help teams incorporate domain knowledge while engaging in theory-driven evaluation practices. For example, HCI, ML, and Visualization researchers might develop toolkits that help teams map hypothesized causal relationships between interventions, outcome variables, and latent constructs during problem formulation. In education, this tool might help teams chart students' longitudinal trajectories through the education system, weigh points of intervention, and consider an intervention's effect on multiple downstream outcomes. This tool might also encourage teams to draw upon their domain knowledge to identify spurious causes of outcome variables --- e.g., by listing causes of an outcome that are unrelated to the construct of interest. During the design probe, several participants suggested an interest in a tool supporting these evaluation practices. While the \textit{DoWhy} package \citep{sharma2020dowhy} enables data scientists to reason about causal relationships in their data, this package is not tailored to support the formulation of prediction tasks.\looseness=-1 

\lgedit{Our findings also demonstrate a need for tools to help teams translate their theoretical understanding of a construct to low-level assertions about expected model behavior. \textit{Behavioral evaluation}, which involves testing the capabilities of a system against a specification of requirements \citep{rahwan2019machine, cabrera2023did},} is one paradigm for facilitating this translation. Prior HCI research has explored leveraging behavioral evaluation to help data scientists evaluate the accuracy and fairness of models \citep{cabrera2023zeno}. \lgedit{Future research might extend this approach to facilitate improved measurement practices. For instance, such a toolkit might enable data scientists to verify the scores assigned by a predictive model against \textit{anchor points}, or cases known to be positive or negative for the construct of interest. When anchor points are unavailable, teams might also check consistency of predictions against a partial ordering over cases --- e.g., student essays sorted by a domain expert for their expected ``authenticity." Supporting both theory- and data-driven activities is critical for helping teams close the gap between their abstract understanding of a construct and its operationalization in a modeling formulation}.\looseness=-1

\lgedit{
\subsection{Implications for Generative AI Evaluation}

While we study measurement in the context of predictive modeling, our findings also connect to emerging discourse surrounding the evaluation of Generative AI systems. Recent work has cast Generative AI evaluation as a measurement task, in which evaluation designers measure properties of AI systems — such as the ``toxicity’’ or ``helpfulness’’ of outputs  — by (1) constructing a systematic definition for a concept, (2) operationalizing this definition into a measurement instrument (e.g., annotation instructions), and (3) applying this operationalization to obtain measurements \citep{wallach2024evaluating}. Recent empirical studies have also shown that the evaluation design process is deeply iterative: practitioners require multiple rounds of refinement to reconcile their conceptual understanding of evaluation criteria with measurements obtained on specific model outputs \citep{he2025prompting, ashktorab2024aligning, szymanski2024comparing, shankar2024validates, pan2024human}. Teams may balance a range of practical criteria while engaging with this process, such as the extensibility and actionability of measurement instruments \citep{harvey2025understanding}. More broadly, viewing Generative AI evaluation design as a bricolage practice opens a space of research questions. How do evaluation designers make do when balancing validity with competing criteria under resource constraints (e.g., limited human ratings and real-world usage data)? How might we develop a scaffolding for this iterative process that enhances validity, while also supporting flexibility? Our work offers an entry point to these key questions.\looseness=-1}

\subsection{Study Limitations}

\lgedit{Our study has several important limitations. First, by foregrounding data scientists' perspectives, we capture only one dimension of a deeply collaborative process. During interviews, data scientists described collaborating with domain experts, such as clinicians or educators, to refine label definitions and evaluate the validity of their formulations. Recent work has also shown that community members provide critical feedback on problem formulations \citep{kuo2023understanding, haque2024we}. As a result, the factors shaping how bricolage unfolds likely extend beyond participants' vantage points as data scientists. 

Our study also characterizes how problem formulation unfolds in a limited range of domains. We interviewed participants working in education and healthcare --- where specific regulatory constraints, data ecosystems, and incentives may have shaped their bricolage practices. Data scientists in domains with more readily available data (e.g., social media platforms), different regulatory environments (e.g., finance), or distinct organizational incentives (e.g., user engagement versus educational assessment) may consider criteria beyond those documented in our findings.

Finally, our findings draw upon data scientists with a baseline level of domain knowledge and technical competency. Our recruitment criteria stipulated that data scientists have at least one year of full-time experience in the healthcare or education sectors to participate. Additionally, our quoted examples of (re)formulation strategies in Section \ref{subsec:strategies} often feature creative practices by highly experienced practitioners. Less experienced data scientists may draw upon a narrower range of strategies in their repertoire or engage in less sophisticated validity evaluation practices. Understanding and supporting novice practitioners' bricolage practices is an important area for future research.\looseness=-1}

\section{Conclusion}

In this study, we interviewed fifteen data scientists working in education and healthcare domains to understand their practices, challenges, and perceived opportunities for target variable construction in predictive modeling. We explore how data scientists adopt a \textit{bricolage} approach to target variable construction, designing operationalizations for unobserved, latent constructs by \textit{making do} with the data that they \textit{already} have at their disposal. We characterize the contours and constraints of data scientists’ bricolage practices, including the problem (re)formulation strategies they apply to balance across different criteria. Critically, we argue that \textit{interventions} designed to improve target variable construction should begin by acknowledging the inherent uncertainty and resource constraints involved in real-world data science work. By understanding and designing to support target variable construction \textit{as a bricolage process}---for example, by helping teams navigate validity tradeoffs, determine when and how to collect more data, and identify appropriate (re)formulation strategies---the research community can help data scientists more thoughtfully engage with problem formulation and avert the negative consequences that arise from poor modeling decisions.

\section{Acknowledgments}

We thank our research participants for making this work possible and anonymous reviewers for their thoughtful feedback. This work was supported by an award from the UL Research Institutes through
the Center for Advancing Safety of Machine Intelligence (CASMI) at Northwestern University and the National Science Foundation Graduate Research Fellowship Program (Award No. DGE1745016).

\bibliographystyle{ACM-Reference-Format}
\bibliography{refs}

\appendix

\newpage
\appendix
\section{Details of the Vignette-Based Problem Formulation Task}\label{appendix:design_probe_details}

We began the design activity by introducing participants to a modeling scenario matched to their area of domain expertise. We asked participants working in healthcare to imagine that they were developing a model to support emergency room triage decisions by predicting patients with high \textit{``medical acuity.''} We told participants that patients' electronic medical records and current symptoms were available as predictors, and described five downstream health outcomes that were available as prediction targets: hospitalization, 3-day ER readmission, in-patient mortality, 12-hour length of stay, and 12-hour ICU admission. To understand how participants assessed the \textit{face validity} of alternative formulations, we provided them with the base rate of each outcome and asked them to reason about which might serve as \textit{``better or worse''} measures of acuity given the stated modeling goals. Data for this vignette was drawn from MIMIC-IV ED \citep{johnson2020mimic, goldberger2000physiobank}, a dataset commonly used to benchmark predictive models targeting the same set of outcome variables (e.g., hospitalization, in-patient mortality, etc. ) in the machine learning for healthcare literature \citep{xie2022benchmarking}. 

We asked participants working in education to imagine that they were developing a model to be used as part of an Early Warning System (EWS) for high school sophomores. We told participants that this system would be used to match students predicted to be ``at-risk'' with a personalized academic success intervention. Following US Department of Education guidelines \citep{doe_ews_2016}, we defined ``at-risk'' students as those failing to achieve basic proficiency in core academic subjects (e.g., reading, math). We told participants that they had access to predictors such as students' academic history, extracurricular involvement, and demographic factors, in addition to four outcomes recorded at the end of their senior year: dropout, senior GPA at a C- or below, major disciplinary event (i.e., suspension or expulsion), and course completion. To understand how participants weighted \textit{face validity}, we provided them with the base rate of each outcome and asked them to reason about which might serve as \textit{``better or worse''} measures of academic risk given the stated modeling goals. We leveraged data from the Educational Longitudinal Study (ELS) \citep{ingels2004education} to construct the vignette. 

After introducing the vignette, we presented a series of model evaluations indicating the performance of predictive models targeting each outcome. After showing each plot, we paused and asked participants to reflect on which outcomes might serve as ``better or worse choices'' for the stated modeling goals. We also asked participants to share any additional information that would be useful for informing their thought process. We provide screenshots for each evaluation plot shown to participants in the pages included below.

We piloted our protocol on one participant to check its flow and comprehensibility. Then, as non-pilot participants engaged with the task over subsequent interviews, we refined the specificity of the scenario description and accompanying evaluation plots to build upon participants' questions and feedback. For example, we later added reports of models' sensitivity and specificity, AU-ROC curves, and kernel density plots after several participants reported interest in this information. Furthermore, while the initial vignette offered general descriptions of modeling intervention (i.e., instructing them to \textit{``decide which patients should receive additional medical resources’’} or \textit{``identify high school sophomores in need of student success interventions’’}), we later increased the specificity of the intervention description (provided in bullet points above) to enable participants to reason about the formulation in greater granularity.

\includepdf[pages=-,scale=0.5]{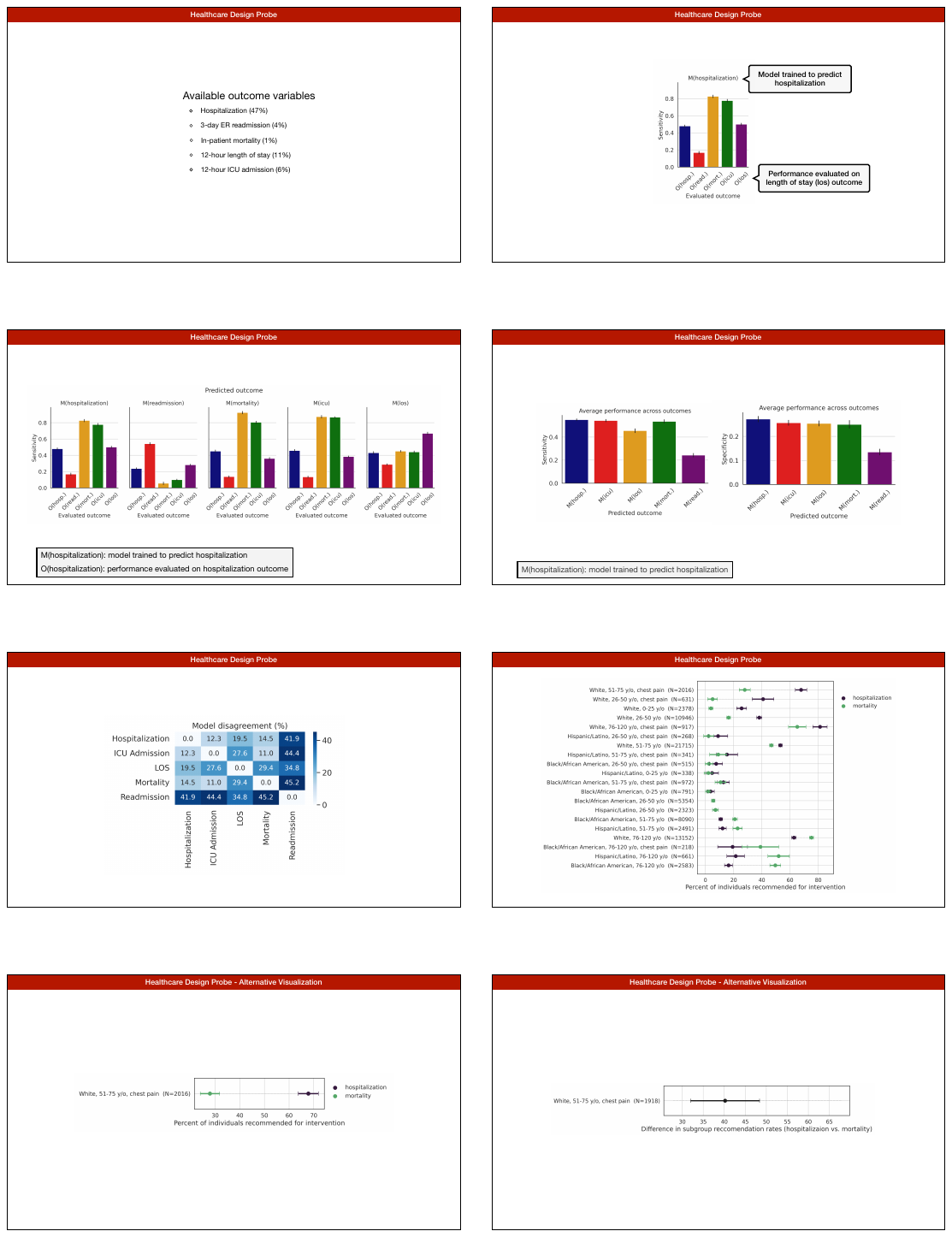}

\end{document}